\documentclass[aps,pra,superscriptaddress,twocolumn]{revtex4-2}

\usepackage{natbib}
\usepackage[colorlinks=true,citecolor=blue,linkcolor=blue,urlcolor=blue]{hyperref}
\usepackage{graphicx}
\usepackage{amsmath,braket} 
\usepackage{amssymb}
\usepackage{color}
\usepackage[normalem]{ulem}
\usepackage[ruled,vlined]{algorithm2e}
\usepackage{changes}
\usepackage{multirow}
\SetArgSty{textnormal}

\renewcommand{\vec}[1]{\mathbf{#1}}
\newcommand{\vtheta}{\boldsymbol{\theta}}

\definecolor{fuchsia}{rgb}{1.0, 0.0, 1.0}

\definecolor{ao}{rgb}{0.0, 0.5, 0.0}

\hypersetup{
    colorlinks=true,
    linkcolor=blue,
    filecolor=magenta,      
    urlcolor=cyan,
}

\begin{document}
\title{Quadratic Unconstrained Binary Optimisation via Quantum-Inspired Annealing}
\author{Joseph Bowles}
\affiliation{ICFO - Institut de Ciencies Fotoniques, The Barcelona Institute of Science and Technology, Castelldefels (Barcelona) 08860, Spain}

\author{Alexandre Dauphin}
\affiliation{ICFO - Institut de Ciencies Fotoniques, The Barcelona Institute of Science and Technology, Castelldefels (Barcelona) 08860, Spain}

\author{Patrick Huembeli}
\affiliation{Institute of Physics, \'Ecole Polytechnique F\'ed\'erale de Lausanne (EPFL), CH-1015 Lausanne, Switzerland}

\author{Jos{\'{e}} Martinez}
\affiliation{Quside Technologies SL, Carrer d'Esteve Terradas, 1, 08860 Castelldefels, Barcelona, Spain}

\author{Antonio Ac\'{i}n}
\affiliation{ICFO - Institut de Ciencies Fotoniques, The Barcelona Institute of Science and Technology, Castelldefels (Barcelona) 08860, Spain}
\affiliation{ICREA - Institucio Catalana de Recerca i Estudis Avan\c cats, Lluis Companys 23, 08010 Barcelona, Spain}

\date{\today}

\begin{abstract}
We present a classical algorithm to find approximate solutions to instances of quadratic unconstrained binary optimisation. The algorithm can be seen as an analogue of quantum annealing under the restriction of a product state space, where the dynamical evolution in quantum annealing is replaced with a gradient-descent based method. This formulation is able to quickly find high-quality solutions to large-scale problem instances, and can naturally be accelerated by dedicated hardware such as graphics processing units. We benchmark our approach for large scale problem instances with tuneable hardness and planted solutions. We find that our algorithm offers a similar performance to current state of the art approaches within a comparably simple gradient-based and non-stochastic setting.
\end{abstract}

\maketitle

\section{Introduction}
Combinatorial optimisation is a class of optimisation problems that has applications in nearly all areas of society. Such problems involve searching for an optimal object amongst an often enormous but finite range of potential candidates, and are notoriously difficult to solve. One type of combinatorial optimisation, called \emph{quadratic unconstrained binary optimisation} (QUBO)~\cite{qubo1,qubo2} involves searching for a bitstring that minimises a quadratic function of its elements. QUBO problems have recently attracted considerable attention, largely because they can be naturally tackled by quantum computers by first mapping the problem to the energy minimisation of a classical Ising model, and then promoting this system to a quantum Ising model~\cite{qqubo1,qqubo2,qqubo3,qqubo4,qqubo5,qqubo6,cim,boettcher2019analysis}. By exploiting phenomena such as superposition and entanglement, the hope is that these quantum algorithms provide faster or higher-quality solutions than their classical counterparts. Much of the focus of showing a quantum advantage versus classical optimisation has consequently been focused around the class of QUBO problems; for example, the D-Wave quantum computer~\cite{dwave,dwave2} exclusively solves this type of optimisation problem. 

At the same time, the interest in QUBO problems has inspired new classical algorithms \cite{sb,sb2,isingalgo1,isingalgo2,isingalgo3,simcim} and corresponding optimization devices \cite{isingmachines1,isingmachines2,isingmachines3,isingmachines4,isingmachines5,isingmachines6}. Since these algorithms can be run on digital logic, they can often handle orders of magnitude more variables than current quantum computers, and as such will serve as valuable classical benchmarks as quantum computers increase in size. Furthermore, since it is known that many hard optimization problems can be mapped to the QUBO setting \cite{mappings1,mappings2,mappings3,mappings4,mappings5}, these classical solvers may also lead to improvements in classical optimization heuristics in general. An important question is thus: what types of classical algorithm are best suited to solving large-scale QUBO problems? A standard approach is to use algorithms such as simulated annealing or population annealing \cite{wang2015comparing}, since these algorithms are naturally suited to the discrete nature of QUBO problems. Although they perform well in general, the fast implementation of these algorithms for large-scale problems is limited by hardware (for example, Fujitsu's digital annealer \cite{digitalannealer} is currently limited to 8192 variables due to limitation of CPU cache), and it is not clear how best to parallelize these algorthims since in their standard versions they rely on sequential parameter updates. A more recent method involves classically simulating the dynamical evolution of a physical system. This is generally done using continuous degrees of freedom (such as position and momentum) whose energy is related to the corresponding Ising Hamiltonian. Two recently introduced methods, Toshiba's simulated bifurcation (SB)~\cite{sb,sb2}, and the simulated coherent Ising machine (SIM-CIM)~\cite{simcim} are of this form. The original SB algorithm~\cite{sb} is based on a simulation of classical nonlinear Hamiltonian system of oscillators, and was later developed into two modified versions of the algorithm~\cite{sb2}. SIM-CIM~\cite{simcim} is a classical simulation of the optical neural network called the coherent ising machine~\cite{cim}. Importantly, both these methods are suited to solving large-scale QUBO problems of tens or even hundreds of thousands of variables, owing to the possibility of GPU acceleration of the computational demanding part of these algorithms. Another interesting approach is to map the QUBO problem to the optimisation of a classical neural network. This has been recently investigated via the use of graph neural networks \cite{schuetz2021combinatorial}, autoregressive neural networks \cite{hibat2021variational}, and neural-network quantum states \cite{gomes2019classical,zhao2020natural}.

In this work we explore an alternative, quantum-inspired classical algorithm for QUBO problems. The algorithm is inspired by quantum annealing \cite{qqubo6,farhi2001quantum}, and is called \emph{local quantum annealing} (LQA). In a standard quantum annealing algorithm, one evolves a multi-qubit quantum state through a Schr{\"{o}}dinger evolution that is generated by a time-dependent Hamiltonian, where the ground state solution of the final Hamiltonian encodes the solution to the problem. In LQA, we use the same Hamiltonian to define a time-dependent cost function that we iteratively optimise via a momentum-accelerated gradient descent based approach. In order to make the optimisation tractable, the cost-function is optimised over a subset of product quantum states that is guaranteed to contain the problem solution. In this way, the system stays in a low energy (product) state throughout the optimisation, which can be seen as an approximation of the annealing process. Since we use a gradient descent approach on the energy landscape, the method is not however equivalent to simulating a type of adiabatic quantum evolution in which the system stays in the global minimum at all times, as is the case for quantum annealing. Surprisingly however, for small systems the method is often able to find the global minimum via the route defined by the gradient descent procedure, and for larger systems gives a method for finding good approximate solutions in very short time. 

We note that this approach is reminiscent but not equivalent to those studied in~\cite{smolin2014classical}, which propose to update the parameters via a dynamical physical evolution defined by the energy of the system. We comment more on this in section~\ref{sec:methods}. Similarly to SB and SIM-CIM, the computationally expensive step in the algorithm corresponds to a matrix-vector multiplication which can be accelerated by dedicated hardware such as GPUs and FPGAs, and so the algorithm is naturally parallelizable and requires approximately the same resources per optimisation step as SB and SIM-CIM. Unlike simulated bifurcation however, our approach is purely gradient based, and unlike SIM-CIM, it does not consume randomness during the optimisation. We note that our approach shares some similarities with the molecular dynamics part of the hybrid quantum annealing algorithm presented in \cite{irie2021hybrid}, where it is possible that the recursive form of the leap-frog algorithm used there plays a similar role to gradient descent in our approach.

We benchmark our algorithm against the three alternative versions of the SB algorithm \cite{sb2} and SIM-CIM \cite{simcim}. Since all these algorithms require the same computational resources per optimisation step this makes for a fair and uncontroversial comparison in terms of the solution quality per optimisation step. We focus most of our benchmarking around recently developed methods of planted solutions \cite{planted_tile,planted_wishart,chook}, which provide constructions of QUBO problems with tuneable hardness whose global optimal solution is known. In our opinion this provides a more complete picture than benchmarks that focus on a single problem, or an arbitrary sets of problems whose hardness is unknown. We find that LQA provides comparable solution quality to SB and SIM-CIM over all problem instances and thus opens an alternative route to large-scale QUBO optimisation via a purely gradient based deterministic algorithm. We also believe that the relative simplicity of our approach paves the way for a number of potential improvements or extensions that we discuss at the end of the article. 

An implementation of our algorithm in PyTorch is available as a Python package; see \url{https://github.com/josephbowles/localquantumannealing}.

\begin{figure}
    \centering
    \includegraphics[width=\columnwidth]{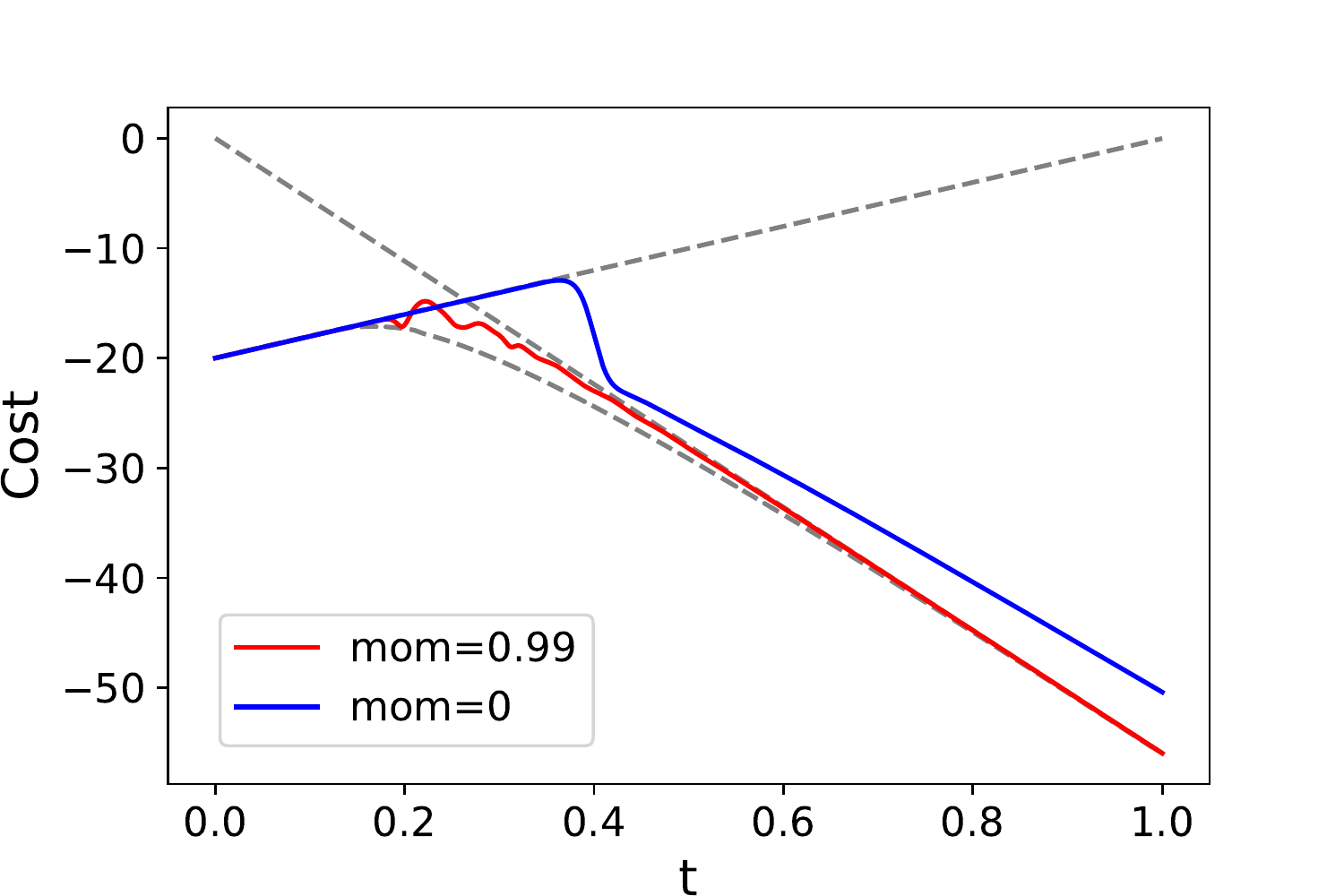}
    \caption{Plot of the cost function value as a function of $t$ for a fully connected 20-spin problem with randomly chosen weights in the interval $[-1,1]$. The dashed straight lines show the cost of the global solutions for the times $t=0$ and $t=1$, evaluated at different values of $t$ (i.e. the energy crossing of the two ground states if the two Hamiltonian $H_x$ and $H_z$ would commute). The lower dashed line shows the minimum of the cost function of the product state ansatz \eqref{product} obtained via an intensive basin-hopping algorithm. Here, we consider a vanilla gradient descent approach (mom=0) and a momentum-assisted approach (mom=0.99). One sees that the use of momentum is particularly effective at helping the system leave the initial minimum, since this point becomes a saddle point for some value of $t$. This helps the momentum assisted approach stay closer to the global minimum throughout the optimisation, and reach the global minimum.}
    \label{fig:methods}
\end{figure}

\section{Methods}\label{sec:methods}
Formally, a QUBO optimisation problem is one of the form 
\begin{align}
    \min_{\vec{x}\in \{0,1\}^n} \quad \vec{x}^T\;Q\;\vec{x} + \vec{x}^T\;\vec{a}
\end{align}
where $Q$ is a $n\times n$ real symmetric matrix such that $Q_{ii}=0 \; \forall i$, and $\vec{a}$ is an $n\times 1$ real vector. By defining the $\pm1$ valued variables $s_i=2x_i-1$, and corresponding vector $\vec{s}$, the problem is equivalent (up to a problem dependent constant) to 
\begin{align}\label{min_classical}
    \min_{\vec{s}\in \{+1,-1\}^n} \quad \vec{s}^T\;J\;\vec{s} + \vec{s}^T\;\vec{b}
\end{align}
where $J=Q/4$, $\vec{b}=(\vec{a}+Q\vec{1})/2$, and $\vec{1}$ is the vector of ones. Note that the second term in \eqref{min_classical} can be incorporated into the first at the expense of adding one extra variable with fixed value $+1$. From hereon we therefore consider problems in the above form with $\vec{b}=0$ so that our problem becomes
\begin{align}\label{min_classical2}
    \min_{\vec{s}\in \{+1,-1\}^n} \quad \vec{s}^T\;J\;\vec{s}.
\end{align}
In this form, the optimisation is equivalent to minimising the energy of a classical Ising Hamiltonian $J$ with no bias field, where the variables $s_i$ are interpreted as classical spin values. 

The minimum \eqref{min_classical2} is equivalently obtained as the ground state of the quantum Ising Hamiltonian 
\begin{align}\label{hamiltonian}
    H_z=\sum_{ij}J_{ij}\sigma_z^{(i)}\sigma_z^{(j)} 
\end{align}
where $\sigma_z^{(i)}$ denotes the Pauli z matrix applied to the $i^\text{th}$ qubit of an $n$ qubit system. That is, one way to find the minimum \eqref{min_classical} is via
\begin{align}\label{min_gen}
    \min_{\ket{\psi}\in \mathbb{C}^{2^n}} \bra{\psi}H_z \ket{\psi}
\end{align}
where $\ket{\psi}$ is a normalised quantum state. Since $H_z$ is diagonal in the z basis, this minimum energy will be attained by one (or more) of the z basis product states. In a quantum annealing algorithm, one attempts to solve \eqref{min_gen} by considering a time-dependent Hamiltonian such as
\begin{align}\label{ham_t}
    H(t) = t H_z \gamma - (1-t)H_x
\end{align}
where 
\begin{align}\label{transverse}
    H_x=\sum_i \sigma_x^{(i)}.
\end{align}
Here $\gamma>0$ controls the relative strength of $H_z$ contribution to the energy. The quantum state of the system is initially prepared in the state $\ket{+}^{\otimes n}$, which is the ground state of $H(0)$, and the system is evolved by changing the Hamiltonian from $t=0$ to $t=1$. If this evolution is done slowly enough, the adiabatic theorem guarantees that the state will stay in the ground state throughout the evolution and the global solution to the problem will therefore be obtained. 

Our algorithm is inspired from the quantum annealing approach. Since a classical simulation of quantum annealing is likely impossible due to the exponential memory requirements needed to store the quantum state vector, we do not consider the set of all quantum states, but restrict ourselves to product states of the form 
\begin{align}\label{product}
    \ket{\vtheta}=\ket{\theta_1}\otimes\ket{\theta_2}\otimes\cdots\otimes\ket{\theta_n}
\end{align}
where 
\begin{align}
    \ket{\theta_i}=\cos\frac{\theta_i}{2}\ket{+}+\sin\frac{\theta_i}{2}\ket{-}.
\end{align}
We therefore have 
\begin{align}\label{min_theta}
    \bra{\theta_i}\sigma_z\ket{\theta_i}=\sin\theta_i\;, \quad\quad \bra{\theta_i}\sigma_x\ket{\theta_i}=\cos\theta_i.
\end{align}
Considering these states, the minimisation \eqref{min_gen} is equivalent to
\begin{align}
    \min_{\vtheta}\bra{\vtheta}H_z \ket{\vtheta}
    &=\min_{\vtheta}\sum_{ij}J_{ij}\sin\theta_i\sin\theta_j 
\end{align}

In analogy with the quantum annealing algorithm we  define a time-dependent cost function corresponding to the energy of the system at time $t$: 
\begin{align}\label{cost1}
\mathcal{C}(t,\vtheta) &= \bra{\vtheta}t H_z \gamma - (1-t)H_x \ket{\vtheta} \nonumber\\
&= t\gamma\sum_{ij}J_{ij}\sin\theta_i\sin\theta_j - (1-t)\sum_i \cos\theta_i
\end{align}
%
In our algorithm, we further parameterise the values $\theta_i$ as $\theta_i=\frac{\pi}{2}\tanh{(w_i)}$ where $w_i\in\mathbb{R}$ so that $w_i\rightarrow \pm\infty \implies \ket{\theta_i}\rightarrow\ket{0},\ket{1}$. The cost \eqref{cost1} therefore becomes
\begin{align}\label{cost}
\mathcal{C}(t,\vec{w}) = t \gamma \;\vec{z}^T J\;\vec{z}\; -(1-t)\vec{x}^T\cdot\vec{1},
\end{align}
with  $\vec{z}=(\sin(\frac{\pi}{2}\tanh w_1),\cdots,\sin(\frac{\pi}{2}\tanh w_n))$,  $\vec{x}=(\cos(\frac{\pi}{2}\tanh w_1),\cdots,\cos(\frac{\pi}{2}\tanh w_n ))$. The gradient of \eqref{cost} with respect to the parameters $\vec{w}$ is 
\begin{align}\label{grad}
    \nabla_\vec{w}\;\mathcal{C}(\vec{w},t) = \frac{\pi}{2} [t\gamma(2 J\vec{z})\circ \vec{x}+(1-t)\vec{z}]\circ a(\vec{w}),
\end{align}
where $\circ$ denotes element-wise multiplication of vectors and $a(\cdot)=1-\tanh^2(\cdot)$ is the derivative of the $\tanh$ function and $a(\cdot)$ acts element-wise.

\begin{figure*}
    \centering
    \begin{minipage}{0.4\textwidth}
    \includegraphics[scale=0.5]{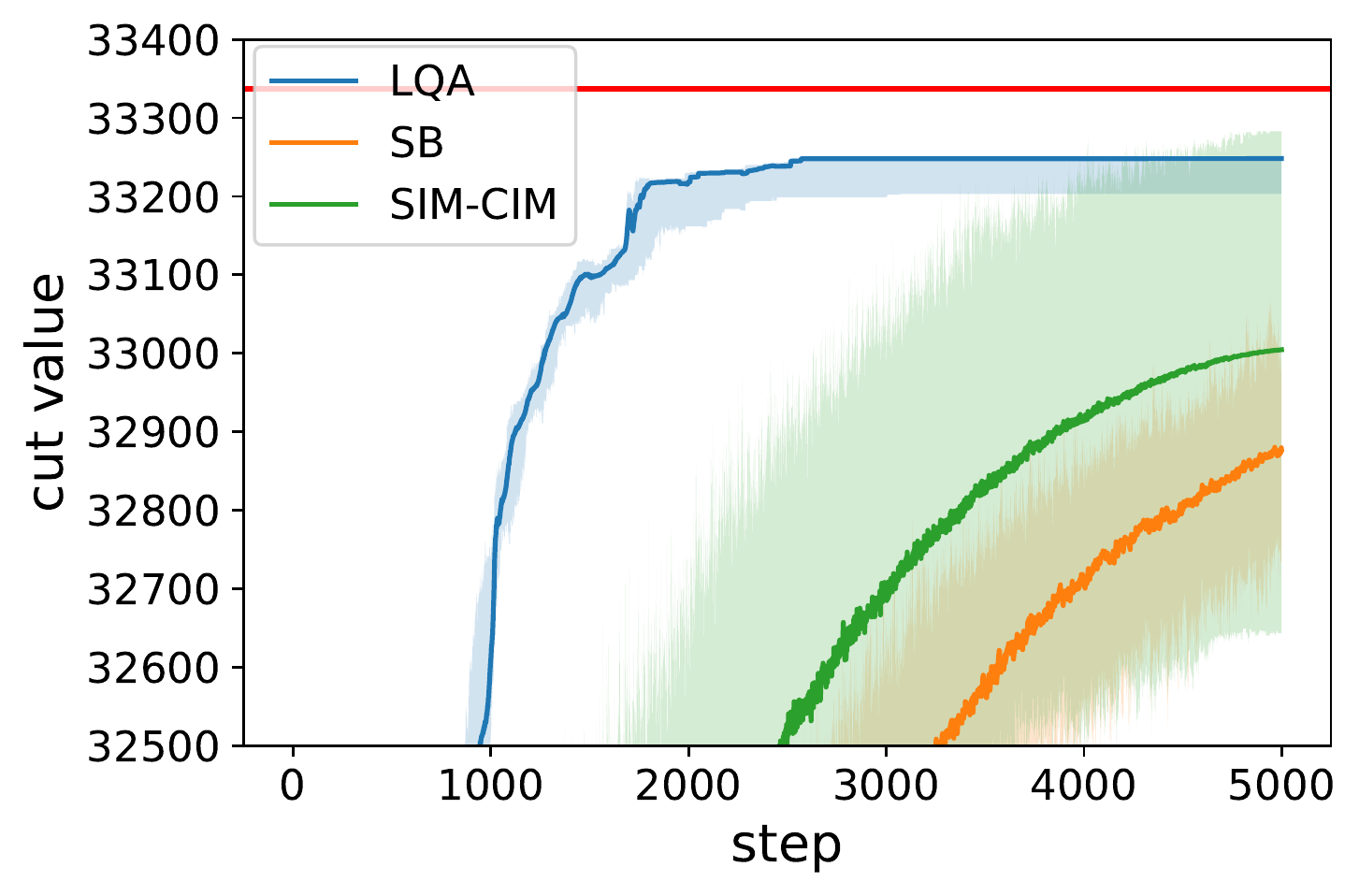}
    \end{minipage}
    \begin{minipage}{0.4\textwidth}
    \includegraphics[scale=0.4]{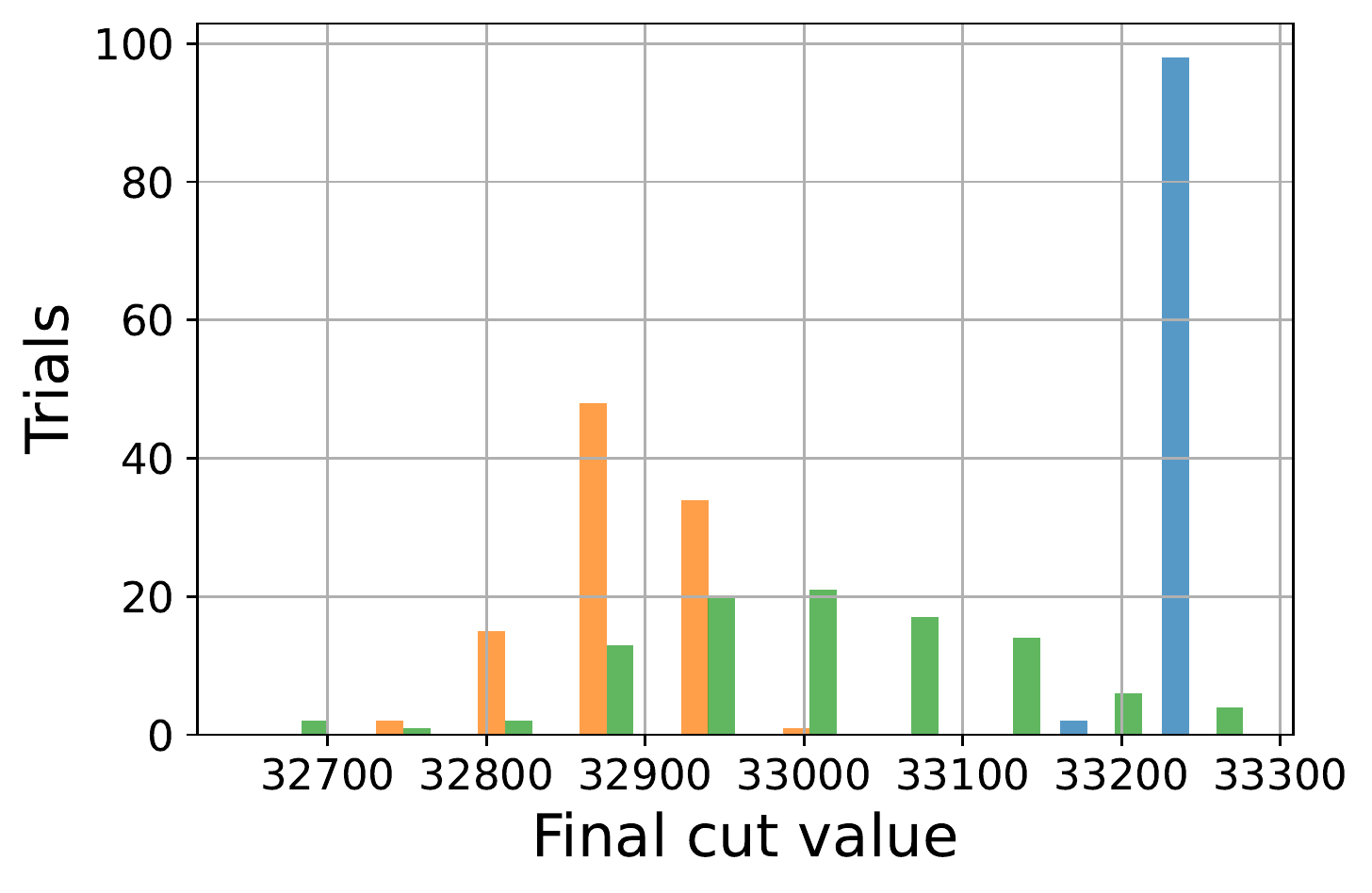}
    \end{minipage}
    \begin{minipage}{0.4\textwidth}
    \includegraphics[scale=0.5]{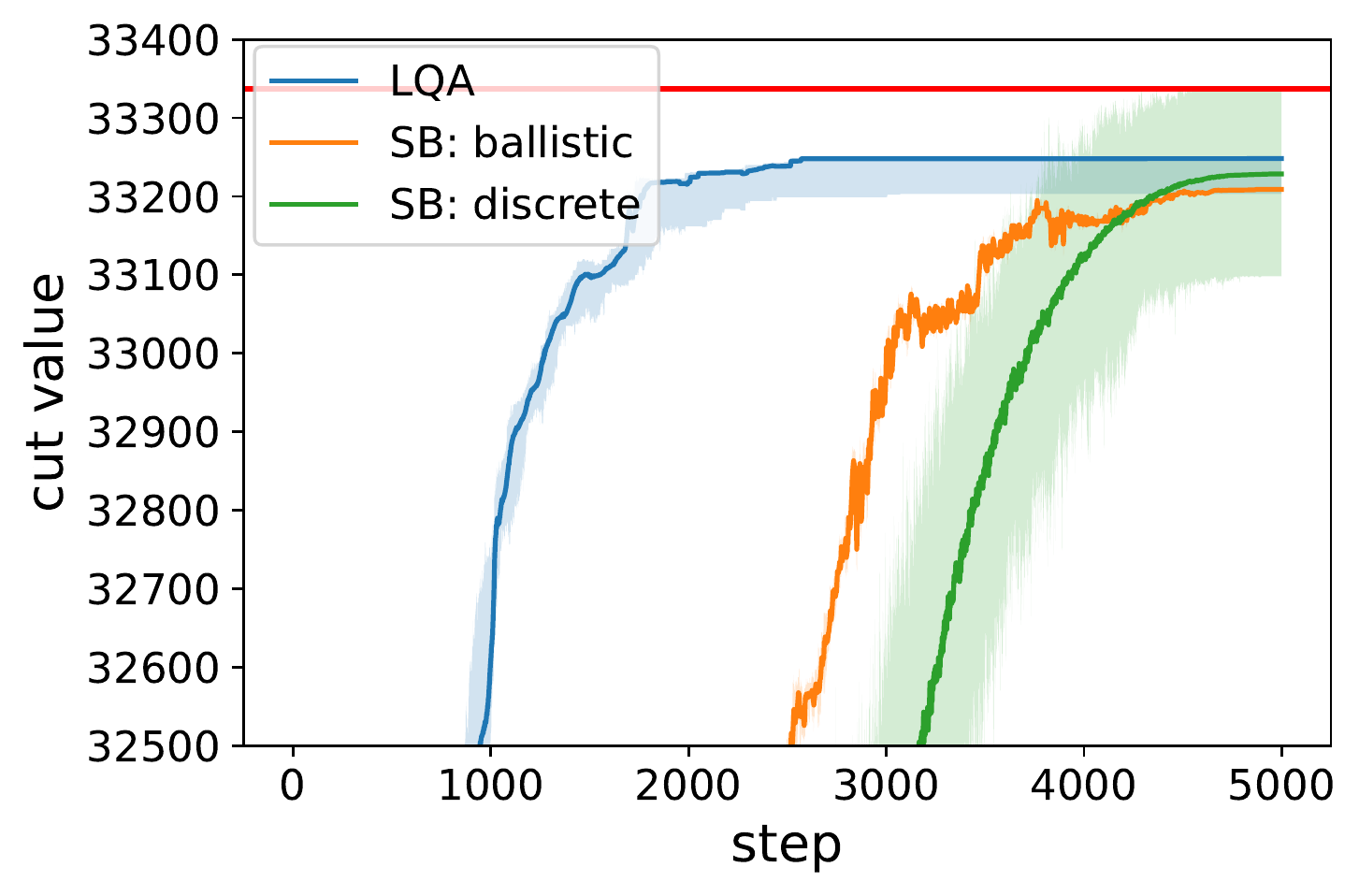}
    \end{minipage}
    \begin{minipage}{0.4\textwidth}
    \includegraphics[scale=0.4]{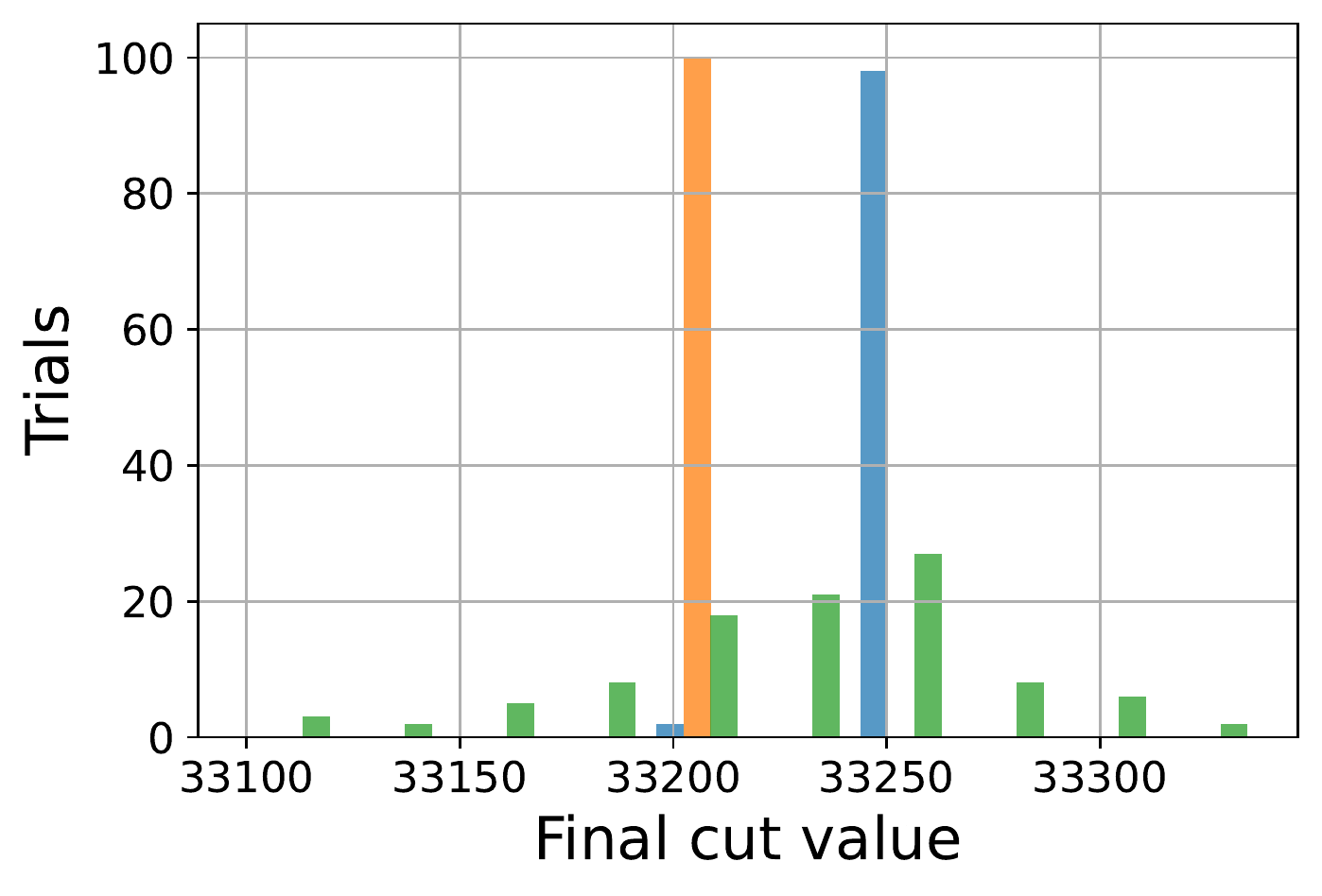}
    \end{minipage}
    \caption{Benchmarking using the $K_{2000}$ Max-Cut problem introduced in \cite{simcim}. This problem is equivalent to solving a 2000 spin fully connected Ising problem with $J_{ij}=\pm1$ chosen uniformly random. Here, we compare LQA to the original simulated bifurcation algorithm \cite{sb} and the simulated coherent Ising mahcine \cite{simcim} (top panel), and to the more recent ballistic and discrete adaptations of the simulated bifurcation algorithm \cite{sb2} (right). We perform 100 trials for each algorithm, each with 5000 optimisation steps. The thick lines show the average values over the 100 trials as a function of the optimisation step. The shaded regions correspond to the max and min cut values over all trials. The histograms plot the final obtained cut values at the end of the optimisation over the 100 trials. For LQA, we chose an initial step size of 1, and set $\gamma=0.1$. For SB, SBB and SBD the step size is set to 0.5, 1.25, 1.25 respectively, with other parameters being set as recommended in \cite{sb2}. For SIM-CIM, a step size of 1 was chosen and the noise parameter $A_n$ set to 0.25. In all algorithms, the initial parameters were set as 0.1*\textsf{rand}, where \textsf{rand} is a random list with elements in the range [-1,1]. \label{fig:k2000}}
\end{figure*}

\begin{figure*}
    \centering
    \begin{minipage}{0.45\textwidth}
    \includegraphics[scale=0.5]{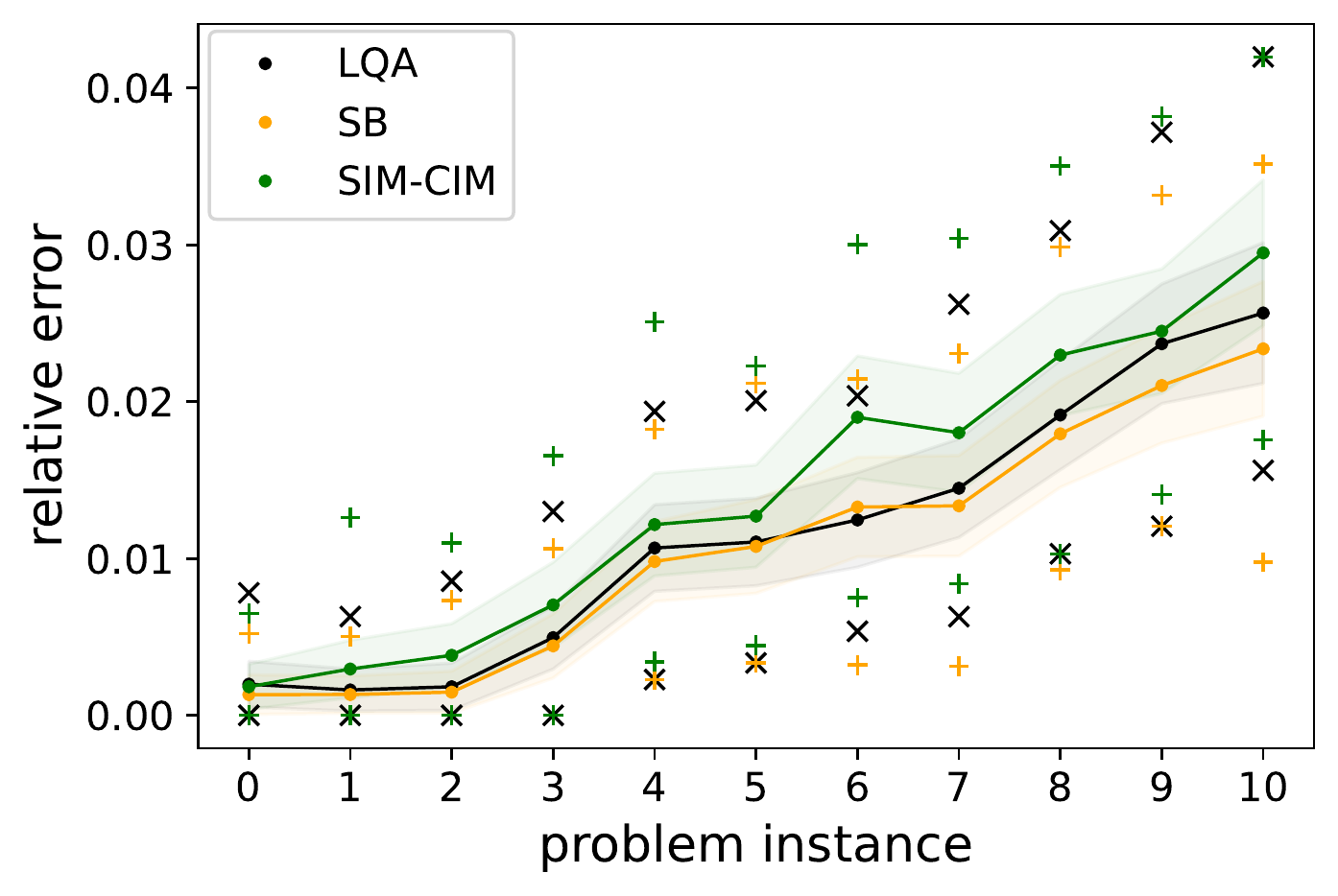}
    \end{minipage}
    \begin{minipage}{0.45\textwidth}
    \includegraphics[scale=0.5]{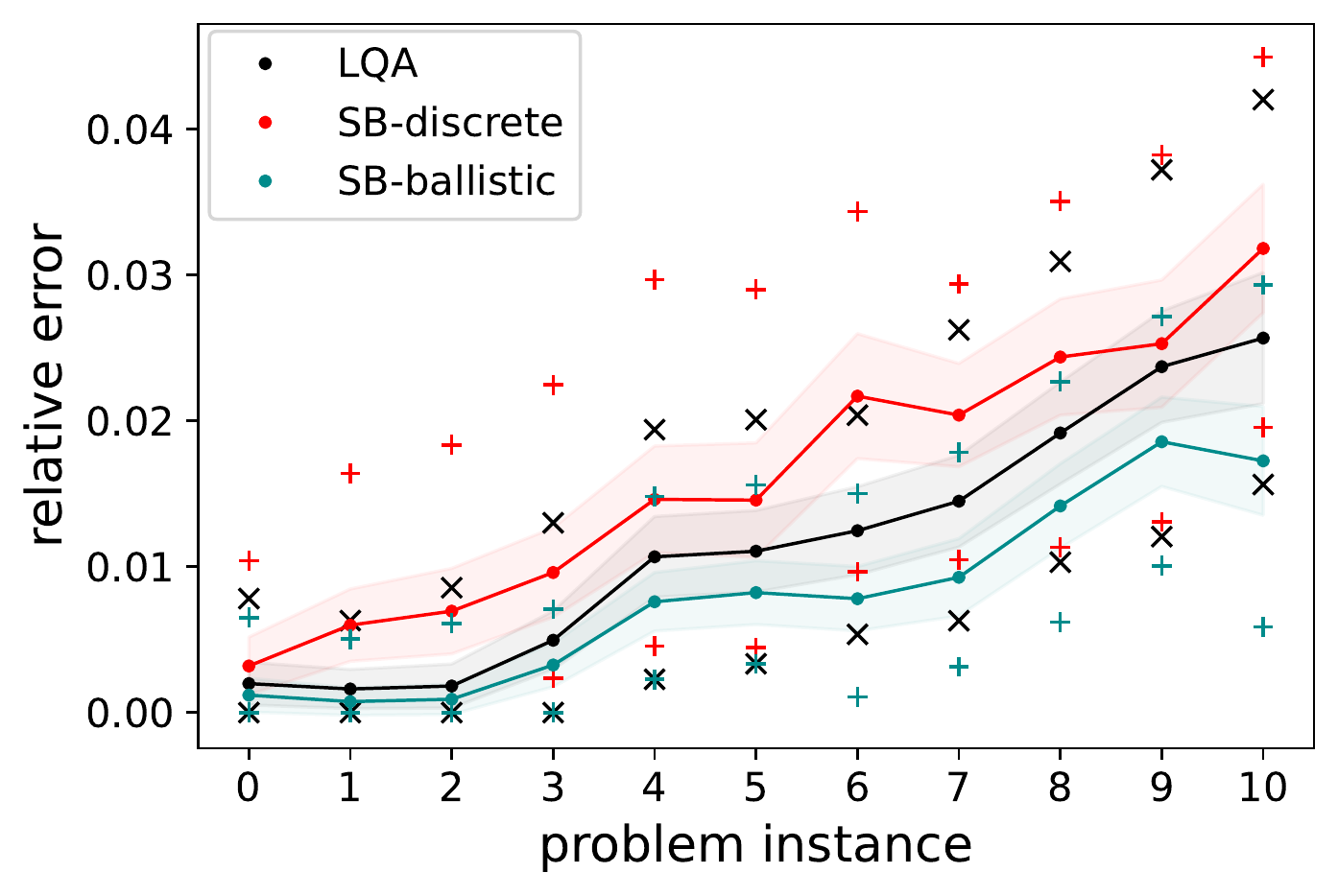}
    \end{minipage}
    \begin{minipage}{0.45\textwidth}
    \includegraphics[scale=0.5]{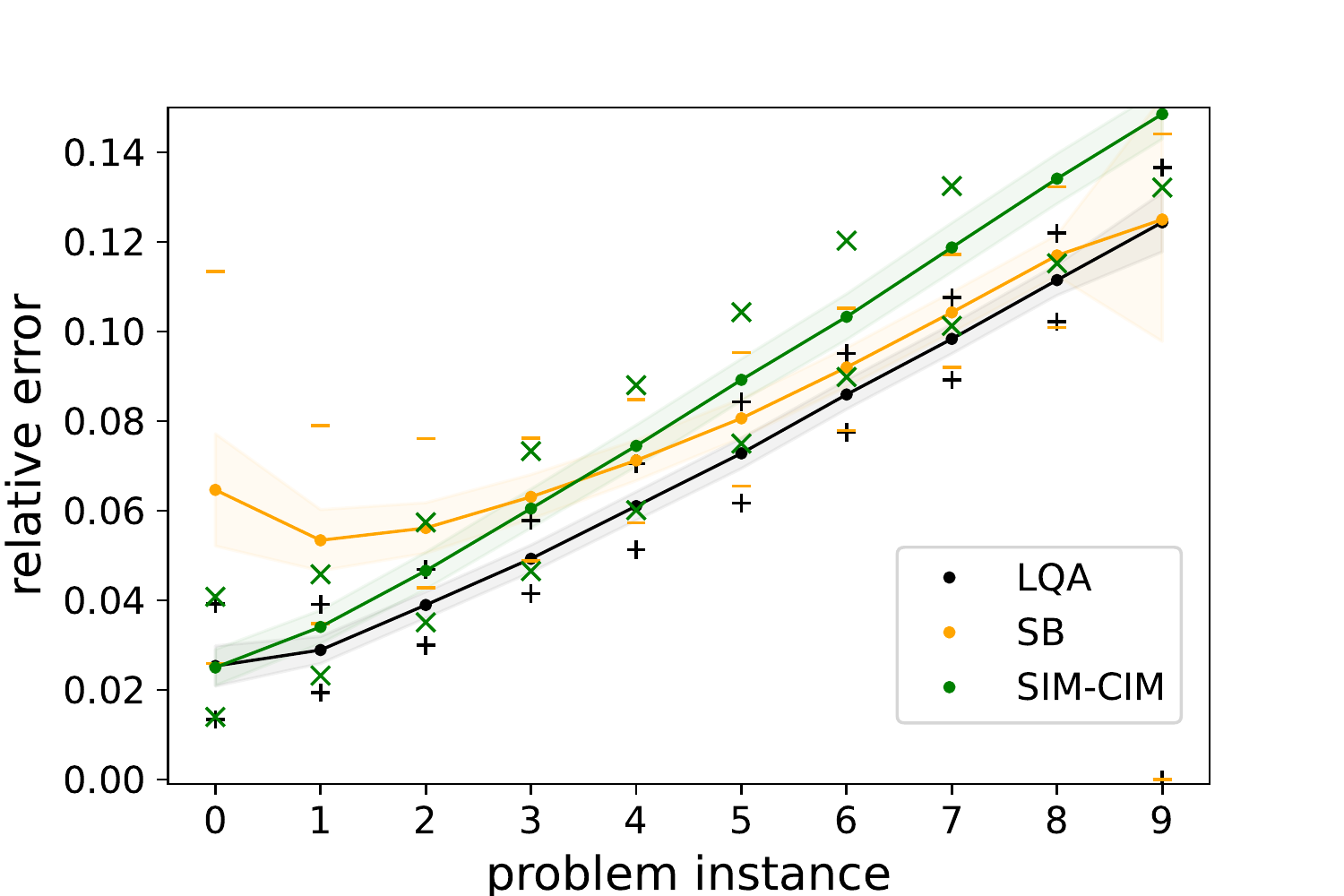}
    \end{minipage}
    \begin{minipage}{0.45\textwidth}
    \includegraphics[scale=0.5]{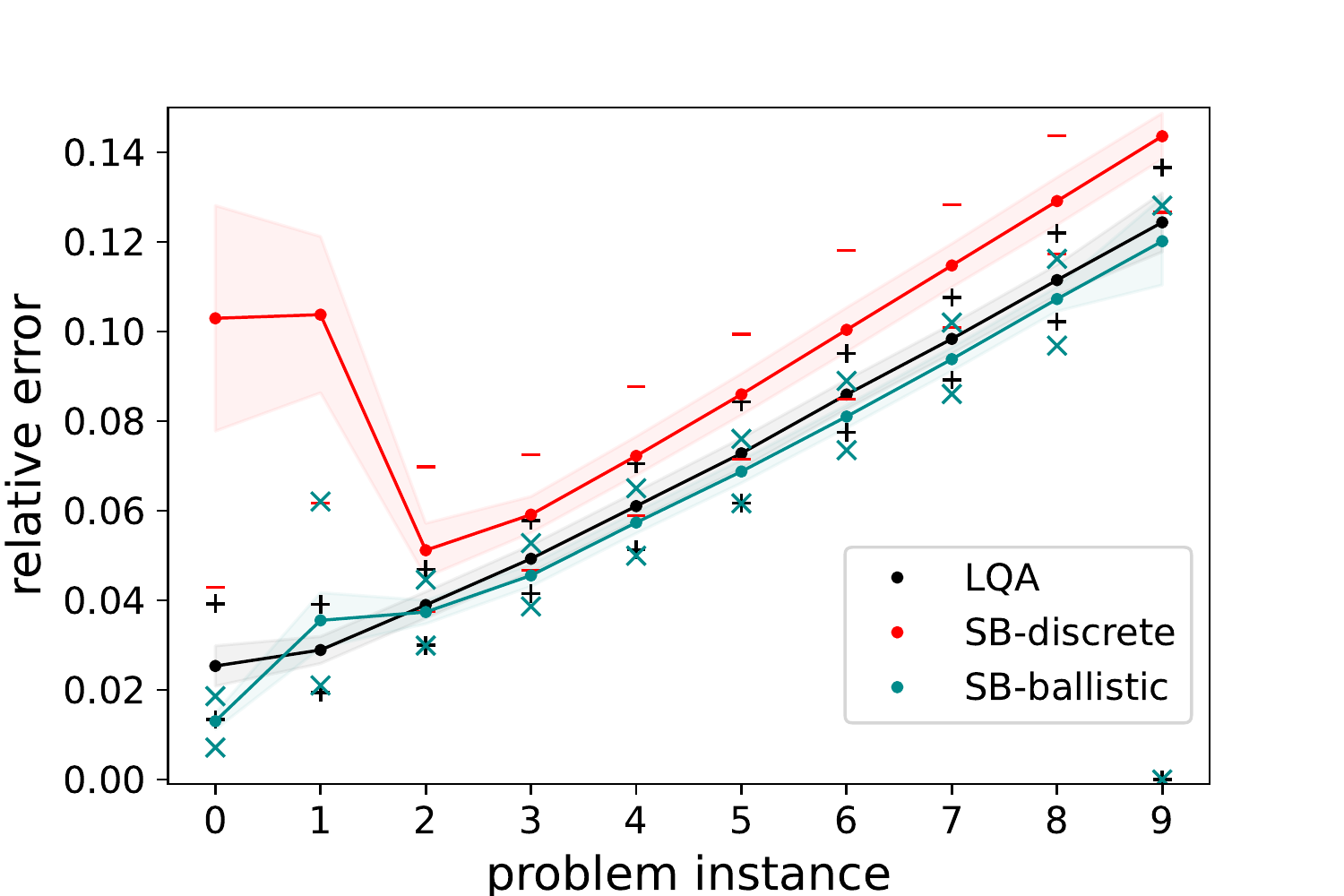}
    \end{minipage}
    \caption{Top panels: Benchmark results for the tile planting problems defined in \cite{planted_tile}, for 11 problem instances (x-axis) with increasing hardness. The number of spins in each problem is 1024. For each instance, 500 trials with 500 steps are performed. The solid lines plot the average relative error to the known global minimum (defined as $\vert\frac{C-C_0}{C_0}\vert$ where $C_0$ is the global minimum and $C$ is the value obtained in a given trial) over the 500 trials. The upper and lower coloured markers denote the maximum and minimum values obtained over the 500 trials. The shaded regions correspond to 1 standard deviation around the mean. Lower panel: Analogous benchmarking results  for the Wishart problems defined in \cite{planted_wishart} which feature and easy-hard-easy problem hardness transition (note that the global minima is obtained for the last instance for some algorithms). Here, 10 problem instance were chosen and 500 trials were performed each with 500 optimisation steps. The initial step size for LQA was set to $2$ for both benchmarks. For SIM-CIM the step size was set to $1$ and $A_n$ to 0.25 for both benchmarks. For tile planting, the step sizes for SB, SBB and SBD were set to $0.5$, $1.25$, $1.25$. For the Wishart planting, the step sizes for SB, SBB and SBD depend on the problem instance due to instabilities in the optimisaiton for the first two problem instances; see appendix A for a table with the precise stepsizes. The parameter initilisation were set as in figure \ref{fig:k2000}.  \label{fig:wishart}}
\end{figure*}

In order to obtain an approximate solution to the QUBO problem, one may use a gradient descent routine on $\mathcal{C}(t,\vec{w})$. More precisely, in each step $i$ of the algorithm, one updates the parameters $\vec{w}$ according to the gradient of $\mathcal{C}(t,\vec{w})$, where $t=i/N$ and $N$ is the total number of steps. A spin configuration can be obtained at any time via $\vec{s}=\text{sign}(\vec{w})$. Here, one may chose from a plethora of strategies that exist for gradient-based optimisation, such as momentum-assistance and adaptive step sizes. For example, the standard momentum-assisted gradient descent technique uses an additional velocity vector $\boldsymbol{\nu}$ (typically initialised as the zero vector) and a momentum parameter $\mu\in[0,1]$, and corresponds to the parameter update (at time t)
\begin{align}\label{momentum}
    &\boldsymbol{\nu} \leftarrow \mu \boldsymbol{\nu} - \eta \nabla \mathcal{C}_\vec{w}(\vec{w},t) \\
    &\vec{w}\leftarrow \vec{w} + \boldsymbol{\nu},
\end{align}
where $\eta$ is the step-size of the gradient descent. Momentum can help to accelerate the gradient descent in flat regions of optimisation, and is widely used in the training of machine learning models. Another common technique is the Adam update~\cite{adam}, which uses the momentum technique and further adapts the step size after each update. In the following pseudocode we denote a generic update as $\vec{w}\leftarrow g(\vec{w},\nabla\mathcal{C}_\vec{w}(\vec{w},t))$, where $g$ may incorporate additional information such as momentum and step size. 

\vspace{5pt}
\begin{algorithm}[H]\label{alg_gd}
\SetAlgoLined
\KwData{$J\in\mathbb{R}^{n\times n}$: symmetric Ising matrix; $\vec{w}_0\in\mathbb{R}^n$: initial weights; $N$: total steps}
 initialization $\vec{w}=\vec{w}_0$\;
 \For{$i=1,\cdots, N$}
 {$\vec{w}\leftarrow  g(\vec{w},\nabla\mathcal{C}_\vec{w}(\vec{w},i/N))$}
 \Return{sign$(\vec{w})$}
 \caption{Local quantum annealing}
\end{algorithm}
\vspace{5pt}

The effect of this algorithm is to continuously push the parameters toward a local minimum of the time-evolving cost function. One may therefore hope to mimic the evolution of a quantum annealing algorithm, in which the system stays in the instantaneous ground state of the time-dependent Hamiltonian throughout the optimisation. For this reason, it is best to initialise the parameters close to zero, since the ground state of $\mathcal{C}(\vec{w},0)$ is given by $\ket{+}^{\otimes n}$.

A couple of important points are in order here. First, note that the computationally expensive part of the algorithm is the matrix multiplication $J\vec{z}$ that appears in the gradient calculation \eqref{grad}. This can be accelerated by dedicated hardware such as GPUs or FPGAs, which for large problems results in a tremendous performance boost with respect to using a CPU. Secondly, we perform only a single parameter update for each value of $t$, rather than waiting for convergence to a local minima before stepping $t$, which requires a much larger optimisation time and typically leads to similar results. Even for large optimisation times, the method is not guaranteed to converge to the globally optimal solution, since the optimisation is performed over a much smaller space of product quantum states, and there is no corresponding guarantee that following a locally optimal minimum throughout the optimisation will lead to the ground state of the final Hamiltonian (due to e.g.\ a first order phase transition). Nevertheless, one may expect that the behaviour approximates the quantum annealing behaviour to some extent. 

In figure \ref{fig:methods} we plot the evolution of the cost, given by $\mathcal{C}(\vec{w},t)$ for a simple problem involving 20 spins. Here we also investigate the use of momentum-assisted gradient descent. Although the system does not stay in the global minimum throughout the optimisation (lower dashed curve), the momentum-assisted approach eventually finds the ground state of the system via a different path. The pure gradient-descent algorithm has difficultly leaving its initial parameters $\ket{+}^{\otimes{n}}$, and more frequently does not converge to the global solution. This can be attributed to the fact that the initial local minimum becomes a saddle point from which it is very slow to escape without momentum assistance. We therefore suspect that momentum-assistance is vital in achieving good performance, which we have found to generally be the case. 

\section{Results}
Here we present a number of benchmarking results. We compare our algorithm against the three variants of simulated bifurcation (the original algorithm (SB) \cite{sb}, ballistic simulated bifurcation (SBB) \cite{sb2}, and discrete simulated bifurcation (SBD) \cite{sb2}), and the simulated coherent Ising machine (SIM-CIM) \cite{simcim}. All algorithms were implemented in pytorch on a standard laptop CPU. For LQA, the Adam gradient descent method \cite{adam} was adopted for the parameter updates.

We focus on three types of problem. The first is the $K_{2000}$ Max-Cut problem (see figure \ref{fig:k2000}). This is a benchmark introduced in \cite{cim} that has been tested on both simulated bifurcation and SIM-CIM which exploits the Max-Cut problem mapping to QUBO (see e.g.\  \cite{mappingslucas}). Here, larger values correspond to higher quality solutions. We find that LQA achieves the highest mean value over 100 optimisation trials of 5000 steps each, with SBD achieving the best value over all trials. 
The distribution of final values varies significantly over the algorithms, and LQA and SBB both feature a small variance with a strong peak in a single solution. We note that this behaviour is sensitive to hyperparameter choice (one can achieve a larger variance at the cost of a lower mean value by adjusting the initial step size) and thus seems not to be a generic feature of LQA. 

Next, we consider performance with respect to problems with planted solutions (see figure \ref{fig:wishart}). To generate these problems we make use of the Chook package \cite{chook}.  We consider two such classes: the first are the 2D `tile planting' problems (figure \ref{fig:wishart} top panels) introduced in \cite{planted_tile}. These problems are defined on a 2D lattice and although in principle solvable in polynomial time due to their planar nature, are often challenging for heuristic solvers. Here we consider 1024 spin problems that are constructed from the C2 and C3 tiles (see \cite{planted_tile} for more details). We generate 11 problem instances of this type, where the probability to use a C3 tile is increased linearly from 0 to 1 over the problem instances. It has been observed that this results in progressively harder problems, which is reproduced in our results. For these problems, the best results were obtained by SBB, with LQA giving similar results to SB. 

We also study the class of Wishart planted solutions (figure \ref{fig:wishart} lower panels) defined in \cite{planted_wishart}, which generate fully connected problems with an easy-hard-easy transition. Here we consider 500 spins problems. We find a very similar performance among all algorithms, with SBB and LQA performing the best in the hard central problem region. SBD performed almost identically to SIM-CIM so we do not show these results here for clarity. SB has some problems with stability for the initial problem instance which we suspect could be remedied with appropriate hyper-parameter tuning. For problem instances at the latter easy tail of the sequence, it is known that the probability for local optimisation methods to find the ground state solution by chance increases significantly \cite{planted_wishart}; the relatively few trials in which the algorithms find the global minimum in these cases may thus be more an indication of luck for this particular batch of trials rather than a feature of the algorithm itself.

\section{Discussion}
Our method is reminiscent of but different to the approaches suggested in \cite{smolin2014classical, hatomura2018shortcuts}, which use a similar product state ansatz and Hamiltonian to mimik the effects of quantum annealing. These approaches are based on a dynamical (physical) evolution of the parameters $\vtheta$ under the Hamiltonian \eqref{ham_t}. In its most basic form, our method uses a simple gradient descent update, which results in a different (un-physical) trajectory of the parameters $\vtheta$. Under the momentum-assisted update \eqref{momentum} and in the continuous-time limit (i.e.\ for infinitesimal step size), it is known that a physical interpretation of our approach is possible \cite{qian1999momentum}: namely, as a dynamical evolution of a system in a viscous medium in which the momentum parameter plays the role of mass. This evolution is not equivalent to that of \cite{smolin2014classical,hatomura2018shortcuts} due to the effects of damping implied by the viscosity.  Furthermore, since our parameter updates are done at the level of the variables $\vec{w}$ and not $\vtheta$, the variables on which this physical evolution is understood are not the same. We have found that these differences can result in significant differences in solution quality for large problems. For the case of more complex parameter updates, such as the Adam \cite{adam} update that we use for our benchmarking, it is not clear if the parameter evolution can be understood physically. In any case we note that, as is typical with gradient descent methods, the solution quality can be quite sensitive to the choice of step size, and it is the case that large step sizes often outperform small step sizes, even for long optimisation times. Given the considerations, our method could also be be viewed as a type of gradient-descent based graduated optimisation \cite{gradopt1,gradopt2,gradopt3,gradopt4} (also called the `continuation method').

The performance of the tested algorithms is similar throughout all benchmarks, with no algorithm clearly outperforming another. This is perhaps not surprising, since although they are all designed from quite different starting points, the parameter updates all make use of the same matrix-vector product between the coupling matrix $J$ and a parameter dependent vector that encodes the solution, which becomes more dominant as the optimisation progresses. For LQA, this is the product $J\vec{z}$ in equation \eqref{grad}. It is therefore unclear whether one should expect any large difference in performance between any of these algorithms since they may all feature basins of attraction to similar solution qualities despite their differing parameter updates. We would argue however, that LQA may be the most versatile of the options. Firstly, it is pure gradient-descent based; it can therefore make use of myriad of tools from machine learning for gradient based optimisation, as well as be used as a subroutine in any other gradient-based algorithm. The use of Adam in our tests makes it quite stable to initial step size variation, which to some extent removes the burden of setting this hyperparameter. Although not done here, second order derivatives of the cost could be calculated analytically via \eqref{grad}. Thus, methods that make use of second order information could be incorporated exactly and may give a performance boost. 

We believe that the main value of the algorithm however is the form of the time-dependent cost function, since this results in significantly better solutions than using a static Hamiltonian. It would be interesting to investigate if this approach could be improved. For example, although we use a simple linear annealing schedule here, non-linear schedules may give better results. On this note, there have been a number of works which use alternative transverse Hamiltonians in order to improve the performance of quantum annealing \cite{alt_anneal1,alt_anneal2}. It would be interesting to investigate if the different cost landscapes implied by these Hamiltonians could lead to improvements. Finally, it would be interesting to see if the product ansatz \eqref{product} could be expanded to include a wider range of quantum states without significantly sacrificing the speed of the algorithm. Here, previous works connecting QUBO optimisation to neural network quantum states~\cite{gomes2019classical,Carleo602} and matrix product state ~\cite{Bauer15} may be valuable.

\emph{Related work}---We note that while preparing this draft we became aware of an independent work \cite{veszeli2021mean} that also suggests using a gradient approach that is similar to ours. 

\emph{Acknowledgments}--- This work was supported by the ICFO-Quside Joint Laboratory in Quantum Processing, Fundacio Cellex, Fundacio Mir-Puig, Generalitat de Catalunya (SGR 1381, SGR 1341, QuantumCAT, and CERCA Programme), ERC AdG NOQIA and CERQUTE, Spanish MINECO (Severo Ochoa CEX2019-000910-S, Plan National FIDEUA  and TRANQI, Retos QuSpin, FPI, QUANTERA MAQS PCI2019-111828-2 / 10.13039/501100011033), EU Horizon 2020 FET-OPEN OPTOLogic (Grant No 899794), and the National Science Centre, Poland (Symfonia Grant No. 2016/20/W/ST4/00314), Marie Sk\l odowska-Curie grant STRETCH No 101029393, a fellowship granted by la Caixa Foundation (ID 100010434, fellowship code LCF/BQ/PR20/11770012), and the AXA Chair in Quantum Information Science.


\bibliographystyle{apsrev4-1_our_style}
\bibliography{references}

\begin{appendix}
\section{parameter setting for Wishart planting benchmark}
The table below displays the step sizes used for the Wishart planting benchmarking of Figure \ref{fig:wishart} for each algorithm. Step sizes were set close to the largest values  possible before instabilities deteriorate the solution quality.\\
\begin{tabular}{ |c|c|c|c|c|c|c|c|c|c|c| } 
\hline
instance:&  0 &  1 & 2 & 3 & 4 & 5 & 6 & 7 & 8 & 9  \\
\hline
LQA & 2 & 2 & 2 & 2 & 2 & 2 & 2 & 2 & 2 & 2 \\ 
SB & 0.5 & 1 & 1 & 1 & 1 & 1 & 1 & 1 & 1 & 1 \\ 
SBB & 1.0 & 1.25 & 1.25 & 1.25 & 1.25 & 1.25 & 1.25 & 1.25 & 1.25 & 1.25 \\
SBD & 0.1 & 0.1 & 1.25 & 1.25 & 1.25 & 1.25 & 1.25 & 1.25 & 1.25 & 1.25 \\
SIMCIM & 1 & 1 & 1 & 1 & 1 & 1 & 1 & 1 & 1 & 1 \\
\hline
\end{tabular}
\end{appendix}

\end{document}